\documentclass[article,showpacs,floats,floatfix,nofootinbib, usenatbib]{mn2e}

\usepackage{amsbsy}
\usepackage{amssymb}
\usepackage{amsmath}
\input{epsf}
\usepackage{graphicx}
\usepackage{color}

\usepackage{amssymb}
\usepackage{bm}
\def\spose#1{\hbox to 0pt{#1\hss}}

\def\lta{\mathrel{\spose{\lower 3pt\hbox{$\mathchar"218$}}
     \raise 2.0pt\hbox{$\mathchar"13C$}}}
\def\gta{\mathrel{\spose{\lower 3pt\hbox{$\mathchar"218$}}
     \raise 2.0pt\hbox{$\mathchar"13E$}}}

\newcommand{\x}{\mathrm{x}}

\newcommand{\e}{\mathrm{e}}

\def\beq{\begin{equation}}
\def\eeq{\end{equation}}
\def\bea{\begin{eqnarray}}
\def\eea{\end{eqnarray}}

\def\x{{\rm x}}

\def\n{{\rm n}}
\def\p{{\rm p}}
\def\e{{\rm e}}

\begin{document}
\title{The g-mode spectrum of reactive neutron star cores}

\author[N. Andersson \& P. Pnigouras]{N. Andersson and P. Pnigouras\\
Mathematical Sciences and STAG Research Centre, University of Southampton, Southampton 
SO17 1BJ, UK}

\date{\today}

\maketitle

\begin{abstract}
We discuss the impact of nuclear reactions on the spectrum of gravity g-modes of a mature neutron star, demonstrating the anticipated disappearance of these modes when the timescale associated with the oscillations is longer than that of nuclear reactions. This is the expected result, but different aspects of the demonstration may be relevant for related problems in neutron star astrophysics. In particular, we develop the framework required for an explicit implementation of finite-time nuclear reactions in neutron star oscillation problems and demonstrate how this formulation connects with the usual bulk viscosity prescription. 
We also discuss  implications of the absence of very high order g-modes for problems of astrophysical relevance.
 \end{abstract}

\section{Introduction}

Mature neutron stars support a family of gravity g-modes, that owe the buoyancy associated with a varying (with density) composition to their existence \citep{rg}. 
Similarly, hot young neutron stars may exhibit g-modes supported by entropy gradients \citep{ferr}. However, the thermal buoyancy weakens as the star cools, leading to a gradual evolution of the g-mode spectrum \citep{chris}. Still, the g-modes of a hot neutron star may impact on gravitational-wave emission from the immediate aftermath of the core collapse in which the star was born \citep{ott}. They may also be relevant for attempts to infer stellar parameters through asteroseismology. The g-modes that remain as the star cools down may also impact on observations. In particular, a number of studies have considered possible resonances between the g-modes and the tide induced by a binary partner, and to what extent resonant mode excitation may affect the gravitational-wave signal from an inspiralling neutron star binary \citep{1995MNRAS.275..301K,1999MNRAS.307.1001L,natide} -- a topical issue following the spectacular GW170817 merger event \citep{170817}. 

The g-modes also enter the tidal problem in a somewhat unexpected way, as the high-order modes may couple nonlinearly to the tide and high-order pressure p-modes, leading to a non-resonant instability \citep{2013ApJ...769..121W,pg2}. While this p-g instability is not yet fully understood, a search for its signature in the GW170817 data has been carried out [albeit with an inconclusive result, see \citet{ligopg}]. Finally, the g-modes are important for the dynamics of isolated neutron stars. In particular, they are thought to be key to the saturation (again through nonlinear mode coupling) of the gravitational-wave driven instability of the fundamental f-mode in fast spinning stars \citep{pnig}, and they may play a similar role in the r-mode instability problem \citep{arras}. Given this, there has been some effort to understand the interplay between buoyancy and inertial effects in  spinning neutron stars \citep{grot1,gk}. In particular, it has been demonstrated that the g-modes become dominated by the inertia at high rotation rates.

In essence, a detailed understanding of the g-mode spectrum of realistic neutron star models may be required for progress on a range of relevant astrophysics problems. Moreover, 
as the g-mode spectrum depends on both the matter composition and state of matter [with the presence of superfluid components being particularly significant, see \citet{1995A&A...303..515L,comer,kantor,pass}], it is interesting to ask to what extent one may be able to use  observations to constrain high-density physics. 
This short note should be viewed in that context. Focussing on composition g-modes of cold neutron stars, we provide an explicit demonstration of the ``disappearance'' of g-modes with oscillation period longer than the timescale of the involved nuclear reactions. This result is not in any way unexpected. Nevertheless, we believe it is a worthwhile discussion. In particular, the (formal) absence of very low frequency g-modes helps explain the absence of tidal resonances in well separated binaries (a problem we will explore elsewhere). A cut-off in the high-order g-mode spectrum may also be relevant for nonlinear mode-coupling scenarios (e.g. relating to the p-g instability or the saturation of unstable modes). Finally, our simple analysis highlights the resonant behaviour associated with bulk viscosity due to nuclear reactions. In fact, the way we set up the problem outlines an alternative to the standard bulk viscosity description, which may be useful for a wider range of problems (including numerical simulations). 

\section{Formulating the problem}

\subsection{Fluid perturbations}

As the main aim of our discussion is to provide a proof of principle, we focus on a relatively simple setting. In order to elaborate on the central question, we need to consider perturbations of a background star while keeping track of deviations from chemical equilibrium induced by the fluid motion. This introduces the relevant reaction timescales into the problem and allows us to consider the behaviour in different limits. 

As we are dealing with local physics, it is natural to use a Lagrangian approach to the perturbation problem \citep{fs78}.
Assuming that the star is non-rotating, we then first of all have the  perturbed continuity equation (for the density $\rho$)
\beq
\partial_t \left( \Delta \rho + \rho \nabla_i  \xi^i \right)  = 0 
\label{continuity}
\eeq
where $\xi^i$ is the Lagrangian displacement vector associated with the perturbation
\beq
\Delta  = \delta + \mathcal L_\xi
\eeq
with $\delta$ the corresponding Eulerian perturbation (and $\mathcal L_\xi$ the Lie derivative along $\xi^i$) such that
\beq
\Delta v^i = 
\delta  v^i = \partial_t  \xi^i
\label{delv}
\eeq
The perturbed Euler equation is then
\beq
\partial_t^2  \xi_i +{1\over \rho}  \nabla_i \delta p - {1\over \rho^2} \delta \rho \nabla_i p + \nabla_i \delta \Phi = 0
\label{eul2}
\eeq
where $p$ is the fluid pressure and $\Phi$ is the gravitational potential.

We also  have the Poisson equation for the perturbed gravitational potential 
\beq
\nabla^2 \delta \Phi = 4\pi G \delta \rho
\label{pois1}
\eeq
while the unperturbed background configuration (assumed to be in hydrostatic equilibrium) is such that 
\beq
\nabla_i p = - \rho \nabla_i \Phi \equiv - \rho g_i 
\label{baro}
\eeq
where we have introduced the gravitational acceleration $g_i$, for later convenience.

\subsection{Adding reactions}

As we have already indicated, 
we are interested in the impact of nuclear reactions on the composition g-modes. As this involves keeping track of the matter composition, we take as our starting point a two-parameter equation of state $p=p(\rho,x_\p)$, where $x_\p=n_\p/n$ is the proton fraction. It is worth noting that, in the Newtonian case considered here, the mass density is simply $\rho = m_B n$ where $n$ is the baryon number density and $m_B$ is the baryon mass. Hence, we can think of $\rho$ as a proxy for the number density. Moreover,  the continuity equation \eqref{continuity} remains unchanged (although, strictly speaking, it now represents baryon number conservation). 

In order to account for nuclear reactions, we first of all introduce a new dependent variable $\beta=\mu_\n-\mu_\p-\mu_\e$ which encodes the deviation from chemical equilibrium (with $\mu_\x$, $\x=\n,\p,\e$ the chemical potentials for neutrons, protons and electrons, respectively). For simplicity, we assume a pure npe-matter neutron star core (cold enough that it is transparent to the neutrinos generated in the reactions), which means that the relevant reaction timescales are those associated with the Urca reactions. We then have
\beq
dp = \sum_\x n_\x d\mu_\x = n ( d\mu_\n - x_\p d\beta) 
\label{dprel1}
\eeq
where we have assumed (local) charge neutrality ($n_\p=n_\e$). As the background configuration is in both hydrostatic and beta-equilibrium we only account for reactions at the level of the perturbations. To do this we, first of all, need to note that \eqref{dprel1} should also hold at the (Lagrangian) perturbative level. That is, we have
\beq
\Delta p =  n ( \Delta\mu_\n - x_\p \Delta\beta) 
\eeq
or, as it turns out to be more convenient to work with $\rho$ and $\beta$,
\begin{multline}
\Delta p = \left( {\partial p \over \partial \rho} \right)_{\beta} \Delta \rho +  \left( {\partial p \over \partial \beta} \right)_{\rho} \Delta \beta 
\\
= c_s^2  \Delta \rho +  \left( {\partial p \over \partial \beta} \right)_{\rho} \Delta \beta 
\label{eqone}
\end{multline}
where $c_s^2$ is sound speed of matter in chemical equilibrium.

The question is, what can we say about reactions?
For the protons, we have (in general)
\beq
 (\partial_t + v^j \nabla_j ) n_\p +n_\p \nabla_i v^i  =  \Gamma
\eeq
with $\Gamma$ the relevant reaction rate. Combining this with overall baryon number conservation
\beq
 (\partial_t + v^j \nabla_j )n + n \nabla_i v^i = 0
\eeq
we get (assuming that protons and neutrons move together)
\beq
 (\partial_t + v^j \nabla_j ) x_\p = {\Gamma\over n} 
\eeq
However, we assume that the reaction rate relates to perturbations. That is, we need 
\beq
\Delta \left[  (\partial_t + v^j \nabla_j ) x_\p \right] =  (\partial_t + v^j \nabla_j )\Delta x_\p  = {\Gamma\over n} 
\label{eqtwo}
\eeq
where, at least for small deviations from equilibrium \citep{hans,reis95},
\beq
\Gamma \approx \gamma \Delta \beta 
\eeq
with the  $\gamma$ coefficient encoding the reaction rates.

Thinking of $\beta$ as a function of $\rho$ and $x_\p$, and assuming that the star is non-rotating (so that $v^i=0$), we  have
\beq
\partial_t \Delta \beta = \left( {\partial \beta \over \partial \rho}\right)_{x_\p}  \partial_t \Delta \rho +  \left( {\partial \beta \over \partial x_\p}\right)_{\rho} \partial_t  \Delta x_\p 
\eeq
which, once we use \eqref{eqtwo}, becomes
\beq
\partial_t \Delta \beta = \left( {\partial \beta \over \partial \rho}\right)_{x_\p}  \partial_t \Delta \rho +  \left( {\partial \beta \over \partial x_\p}\right)_{\rho} {\gamma \over n} \Delta \beta  
\eeq
That is, we have
\beq
\partial_t \Delta \beta -  \mathcal A  \Delta \beta = \mathcal B \partial_t \Delta \rho
\label{eqfour}
\eeq
with
\beq
\mathcal A =  \left( {\partial \beta \over \partial x_\p}\right)_{\rho} {\gamma \over n} \ , \qquad 
\mathcal B = \left( {\partial \beta \over \partial \rho}\right)_{x_\p}
\eeq

The coefficients $\mathcal A$ and $\mathcal B$ should be time independent, so if we work in the frequency domain (essentially assuming a time-dependence $e^{i\omega t}$ for the perturbations, without introducing specific notation for the Fourier amplitudes) then we have
\beq
\Delta \beta = {\mathcal B \over 1 + i \mathcal A/\omega} \Delta \rho
\label{dbeta1}
\eeq

Let us now consider the timescales involved. Introducing a characteristic reaction time as 
\beq
t_R = {1 \over \mathcal A} 
\eeq
(noting that the actual timescale is the absolute value of this)
we see that, if the reactions are fast compared to the dynamics (on a timescale $\sim 1/\omega$) then $|t_R \omega| \ll 1$ and we have
\beq
\Delta \beta \approx 0 \ .
\eeq
Basically, the fluid remains in beta-equilibrium. 

However,  in the limit of slow reactions we have $|t_R \omega| \gg 1$ and we can Taylor expand \eqref{dbeta1} to get
\beq
\Delta \beta \approx \mathcal B \left(  1-  i \mathcal A/\omega \right)  \Delta \rho \approx \mathcal B \Delta \rho
\label{dbeta}
\eeq

Using this result in \eqref{eqone}, we have
\beq
\Delta p = \left[ \left( {\partial p \over \partial \rho} \right)_{\beta} + \left( {\partial p \over \partial \beta} \right)_{\rho} \left( {\partial \beta \over \partial \rho}\right)_{x_\p} \right]  \Delta \rho 
\equiv \mathcal C \Delta \rho
\label{eqfive}
\eeq
which leads to
\beq
\delta p = \mathcal C \delta \rho + \left[ \mathcal C \xi^j \nabla_j \rho - \xi^j \nabla_j p \right]
\eeq
However, since
\beq
\nabla_j p = \left( {\partial p \over \partial \rho} \right)_{\beta} \nabla_j \rho = c_s^2 \nabla_j \rho
\eeq
we are left with
\beq
\delta p = \mathcal C \delta \rho +  \left( {\partial p \over \partial \beta} \right)_{\rho} \left( {\partial \beta \over \partial \rho} \right)_{x_\p}  \xi^j \nabla_j \rho 
\label{dprel}
\eeq
The composition of matter impacts on both terms on the right-hand side of this relation. 

Given an actual supranuclear equation of state, the different thermodynamical derivatives required to make the relation \eqref{dprel} explicit should be calculable [although we obviously need to start from a model that does not assume chemical equilibrium from the outset, see for example \citet{pass}].

Before we proceed, let us confirm  that the relations for the slow-reaction case correspond to frozen composition. This serves as a useful ``sanity check'' as it ensures that the mathematics agrees with intuition. We have (first thinking of the equation of state as $p=p(\rho,\beta)$ and then changing to $p=p(\rho,x_\p)$)
\begin{multline}
\Delta p = \left( {\partial p \over \partial \rho} \right)_{\beta} \Delta \rho + \left( {\partial p \over \partial \beta} \right)_{\rho} \Delta \beta \\
= \left( {\partial p \over \partial \rho} \right)_{\beta} \Delta \rho +  \left( {\partial p \over \partial \beta} \right)_{\rho} \left[ \left( {\partial \beta \over \partial \rho} \right)_{x_\p} \Delta \rho +  \left( {\partial \beta \over \partial \x_\p} \right)_{\rho} \Delta x_\p \right] \\
= \left[ \left( {\partial p \over \partial \rho} \right)_{\beta} + \left( {\partial p \over \partial \beta} \right)_{\rho} \left( {\partial \beta \over \partial \rho} \right)_{x_\p} \right]  \Delta \rho\\
+ \left( {\partial p \over \partial \beta} \right)_{\rho}  \left( {\partial \beta \over \partial \x_\p} \right)_{\rho} \Delta x_\p \\
= \mathcal C \Delta \rho +  \left( {\partial p \over \partial \x_\p} \right)_{\rho} \Delta x_\p
\end{multline}
which means that,  if \eqref{eqfive} holds we must have $\Delta x_\p=0$. We also see that
\beq
\left( {\partial p \over \partial \rho} \right)_{x_\p}  = \left( {\partial p \over \partial \rho} \right)_{\beta} + \left( {\partial p \over \partial \beta} \right)_{\rho} \left( {\partial \beta \over \partial \rho} \right)_{x_\p}  = \mathcal C
\label{Cdef}
\eeq
This relation will prove useful later.

\section{The g-mode(s)}

Let us focus on a ``toy version'' of the g-mode problem. Starting from \eqref{eul2}, i.e.
\beq
-\omega^2  \xi_i +{1\over \rho}  \nabla_i \delta p - {1\over \rho^2} \delta \rho \nabla_i p + \nabla_i \delta \Phi = 0
\label{eul3}
\eeq
we first of all make the Cowling approximation (neglect the perturbed gravitational potential $\delta \Phi$), to get
%\beq
%-\omega^2  \xi_i +{1\over \rho}  \nabla_i \delta p - {1\over \rho^2} \delta \rho \nabla_i p  = 0
%\label{eul3}
%\eeq
%or
\beq
-\omega^2  \xi_i +{1\over \rho} \left(  \nabla_i \delta p +g_i  \delta \rho \right)  = 0
\label{eul4}
\eeq

Next we assume a plane-wave solution such that $\delta p \sim e^{ik_ix^i}$ (and similar for all other variables). This leaves us with 
\beq
-\omega^2  \xi_i +{1\over \rho} \left( i k_i \delta p +g_i  \delta \rho \right)  = 0
\label{eul4}
\eeq

\subsection{Fast reactions}

Now let us consider two limiting cases.  First, for fast reactions we have seen that 
\beq
\Delta \beta = 0 \longrightarrow \Delta p = c_s^2 \Delta \rho
\eeq
or
\beq
\delta p = c_s^2 \delta \rho + \xi^j \left( c_s^2 \nabla_j \rho - \nabla_j p \right) =  c_s^2 \delta \rho
\eeq
That is, we have
\beq
-\omega^2  \xi_i +{1\over \rho} \left( i c_s^2 k_i  +g_i \right)  \delta \rho = 0
\label{eul5}
\eeq

However, the continuity equations leads to
\begin{multline}
\delta \rho = - \xi^j \nabla_j \rho - \rho \nabla_j \xi^j  \\
= - {1\over c_s^2} \xi^j \nabla_j p - i \rho k_j \xi^j  =
{\rho \over c_s^2} g_j \xi^j - i \rho k_j \xi^j 
\end{multline}
so we have
\beq
-\omega^2  \xi_i + \left( i c_s^2 k_i  +g_i \right) \left( {1 \over c_s^2} g_j -  i  k_j \right) \xi^j   = 0
\label{eul6}
\eeq
Contracting with the wave vector, we get
\beq
-\omega^2 (k_i \xi^i)  + \left[ i c_s^2 k^2  + (g_i k^i) \right] \left[ {1 \over c_s^2} (g_j \xi^j)-  i  (k_j \xi^j) \right]   = 0
\label{scal1}
\eeq
It is useful to simplify this by assuming short wavelengths. Taking  $k\gg g/c_s^2$ we get
%\beq
%-\omega^2 (k_i \xi^i)  + \left[ i c_s^2 k^2  + (g_i k^i) \right] \left[ {1 \over c_s^2} (g_j \xi^j)-  i  (k_j \xi^j) \right]   = 0
%\label{scal2}
%\eeq
%This leaves us with
\beq
\left[ \omega^2  - c_s^2 k^2 \right] (k_j \xi^j)    \approx 0
\label{scal3}
\eeq
That is, as long as $k_j \xi^j \neq 0$, we have the sound waves
\beq
\omega \approx \pm c_s k
\eeq
The transverse solution, $k_j \xi^j=0$, is trivial. There are no g-modes in this case.

\subsection{Slow reactions}

The case of slow reactions is a little bit more involved. Starting  from
\beq
\Delta p = \left({\partial p \over \partial \rho} \right)_{x_\p} \Delta \rho = \mathcal C  \Delta \rho
\eeq
we find that
\begin{multline}
\delta \rho = {1\over  \mathcal C}  \delta p + \xi^j \left[ {1\over  \mathcal C}\nabla_j p -\nabla_j \rho  \right] \\ 
= {1\over  \mathcal C} \delta p - \rho (g_j \xi^j) \left[ {1\over  \mathcal C} - {1\over c_s^2} \right] 
= {1\over  \mathcal C} \delta p + \rho (A_j \xi^j) 
\end{multline}
where we have introduced
\beq
A_i =   \left[{1\over c_s^2}-{1\over  \mathcal C} \right]  g_i 
\eeq
The radial component
\beq
A = - g \left[{1\over c_s^2}-{1\over  \mathcal C} \right]
\eeq
defines the Schwarzschild discriminant. We also need to use the continuity equation, to get
\begin{multline}
\delta p =  \mathcal C \Delta \rho - \xi^j  \nabla_j p  \\
= - \rho \mathcal C \nabla_j \xi^j  - \xi^j  \nabla_j p 
=  - \rho  \mathcal C \nabla_j \xi^j  + \rho (g_j \xi^j )
 \end{multline}
In terms of plane waves, this means that
\beq
\delta p 
=  - i \rho \mathcal C (k_j \xi^j)  + \rho (g_j \xi^j )
 \eeq

 Moving on to the Euler equation, we still have \eqref{eul5} but this  now leads to
\begin{multline}
-\omega^2  \xi_i + \left[  i k_i + {1\over  \mathcal C} g_i  \right] \left[ - i  \mathcal C (k_j \xi^j)  + (g_j \xi^j ) \right] \\
+ g_i (A_j \xi^j)  = 0
\label{neul4}
\end{multline}

We create two scalar equations by contracting with $g^i$ and $k^i$, respectively. This leads to
\begin{multline}
\left\{ - \omega^2 - i \mathcal C \left[ i k^2 + {1\over \mathcal C} (g_j k^j) \right] \right\} (k_j \xi^j) \\
+ \left[ ik^2 + {1\over c_s^2} (g_j k^j) \right] (g_j \xi^j) = 0 
\end{multline}
and
\begin{multline}
-i \mathcal C \left[ i (k_j g^j) + {1\over \mathcal C} g^2\right] (k_j \xi^j ) \\
+ \left[ -\omega^2 + {1\over c_s^2}  g^2 + i (k_j g^j) \right] (g_j \xi^j) = 0 
\end{multline}
In order for this system to have solutions, we must have
\begin{multline}
\left\{  \omega^2 - \mathcal C \left[  k^2  - {i\over \mathcal C} (g_j k^j) \right] \right\} \left[ \omega^2 - {1\over c_s^2}  g^2 - i (k_j g^j) \right] \\
-  \left[ k^2 - {i\over c_s^2} (g_j k^j) \right]  \left[ g^2 + i \mathcal C  (k_j g^j)\right] = 0 
\end{multline}
Let us simplify this by assuming that $g_j k^j =0$ (which does not change the qualitative nature of the solutions). Then we have
%\beq
%\left(\omega^2 - \mathcal C k^2  \right) \left( \omega^2 - {1\over c_s^2}  g^2 \right) 
%- k^2 g^2 = 0 
%\eeq
%or
\beq
\omega^4 - \left( \mathcal C k^2 + {1\over c_s^2} g^2 \right) \omega^2 - \mathcal C k^2 g A = 0 
\label{quartic}
\eeq
For short wavelengths (as before) this simplifies to
\beq
\omega^4 -  \mathcal C k^2 \omega^2 - \mathcal C k^2 g A = 0 
\label{neul5}
\eeq
and we have two sets of modes. For high frequencies, we get the sound waves
\beq
\omega^2 \approx  \mathcal C k^2 
\eeq
where it is worth noting that the matter composition has a (likely small, but nevertheless) effect on the mode frequency. Meanwhile, the low-frequency modes are given by 
\beq
\omega^2 = -g A = N^2
\eeq
where $N$ is the usual Brunt-V\"ais\"al\"a frequency.
This solution represents the composition g-modes \citep{rg}.

\subsection{The general case}

Suppose we now want to account for finite reaction times. Then the calculation from the previous sections goes through -- pretty much unchanged -- up to equation \eqref{neul5}. However, as we want to make use the general result \eqref{dbeta1},  we need to replace $\mathcal C$ by 
\beq
\mathcal D =  c_s^2+  {\omega \mathcal B \over \omega + i \mathcal A} \left( {\partial p \over \partial \beta} \right)_\rho
\eeq
In order to do this, we need additional thermodynamical derivatives. However, we can use \eqref{Cdef} to get
\beq
\mathcal B \left( {\partial p \over \partial \beta} \right)_{\rho}  = \mathcal C - c_s^2
\eeq
That is, we have
\beq
\mathcal D =  c_s^2+  {\omega \over \omega + i \mathcal A} \left(  \mathcal C - c_s^2 \right)
\eeq
Introducing the reaction time, $t_R=1/\mathcal A$ (as before) , we have
\beq
\mathcal D =  
%c_s^2+  {\omega t_R  \over \omega t_R + i } \left(  \mathcal C - c_s^2 \right)  
%=  {1\over  \omega t_R + i } \left[ ( \omega t_R + i  ) c_s^2+  \omega t_R  \left(  \mathcal C - c_s^2 \right)  \right]\\
%=  {1\over  \omega t_R + i } \left[  \omega t_R   \mathcal C +  i  c_s^2   \right] 
{1 \over  \omega t_R + i } \mathcal C \left[  \omega t_R +  i  {c_s^2 \over \mathcal C}  \right] 
\eeq
so we need to solve
\beq
\omega^4 -  \mathcal D k^2 \omega^2 - \mathcal D k^2 g A = 0 
\label{neul6}
\eeq
with $A$ now  given by
\beq
A  = -g \left[ {1\over c_s^2} - {1\over \mathcal D}\right]
%= {1\over c_s^2}  - {1\over \mathcal C} { \omega t_R + i \over  \omega t_R +  i  {c_s^2 / \mathcal C}}
\eeq
It is also useful to note that
\beq
\mathcal D k^2 g A = 
%k^2 g^2 \mathcal D\left[ {1\over c_s^2} - {1\over \mathcal D}\right]  = k^2 g^2\left[ { \mathcal D\over c_s^2} - 1\right]\\
%= k^2 g^2\left[1 -  {\omega t_R  \over \omega t_R + i }  \left( {\mathcal C \over c_s^2} - 1 \right)  -1 \right] 
%= k^2 g^2 {\omega t_R  \over \omega t_R + i }  \left( {\mathcal C \over c_s^2} - 1 \right) \\
 -  k^2 \mathcal C g^2 {\omega t_R  \over \omega t_R + i }  \left( {1 \over c_s^2} - {1 \over \mathcal C} \right)
\eeq

While it would  be straightforward to solve the problem for a specific stellar model, we prefer to illustrate the involved principles in terms of a suitably simply model problem. Thus we 
parameterise in terms of the mode frequencies in the slow reaction limit:
\beq
\omega^2_f =  \mathcal C k^2 
\eeq
and
\beq
\omega^2_g = g^2   \left[{1\over c_s^2}-{1\over  \mathcal C} \right]
\eeq
Note that, as we can vary $g$ here, the link to $\mathcal C$ and $c_s^2$ is not immediate. However, we know that we must have $\mathcal C> c_s^2$ in order to have oscillatory modes, and we can easily make sure that our model satisfies this constraint. We will also use the non-stratified result
\beq
\omega_0^2 = c_s^2 k^2 
\eeq

In terms of these parameters, we have
\beq
\mathcal D k^2 =  {1 \over  \omega t_R + i } \omega_f^2 \left[  \omega t_R +  i  {\omega_0^2 \over \omega_f^2}  \right] 
\eeq
and
\beq
\mathcal D k^2 g A = - {\omega t_R  \over \omega t_R + i } \omega_f^2\omega_g^2
\eeq
Finally, let us scale the frequencies to the acoustic mode, by introducing $x=\omega/\omega_f$,  $\tilde \omega_0 = \omega_0/\omega_f$ and $\tilde \omega_g = \omega_g/\omega_f$. Similarly parameterising the reaction time (through its relation  to the g-mode frequency)
\beq
t_R =  {\alpha \over \omega_g} \longrightarrow \omega t_R =   {\alpha x \over \tilde \omega_g} \quad \mbox{with} \quad \alpha < 0
\eeq
(where the sign of $\alpha$ ensures that the modes of the toy problem are damped), we arrive at  the final polynomial
%\beq
%{\mathcal D k^2 \over  \omega_f^2} = {\tilde \omega_g \over \alpha x + i  \tilde \omega_g} \left[   {\alpha x \over \tilde \omega_g} +  i \tilde \omega_0^2 \right] 
%\eeq
%\beq
%{\mathcal D k^2 g A \over \omega_f^4}= {\alpha x  \over \alpha x + i  \tilde \omega_g } \tilde \omega_g^2
%\eeq
\beq
x^4 -  {\tilde \omega_g \over \alpha x + i  \tilde \omega_g} \left[   {\alpha x \over \tilde \omega_g} +  i  \tilde \omega_0^2   \right] x^2
 +  {\alpha x  \over \alpha x + i  \tilde \omega_g } \tilde \omega_g^2= 0
 \label{finpol} 
\eeq

\begin{figure}
\begin{center}
\includegraphics[width=0.45\textwidth,clip]{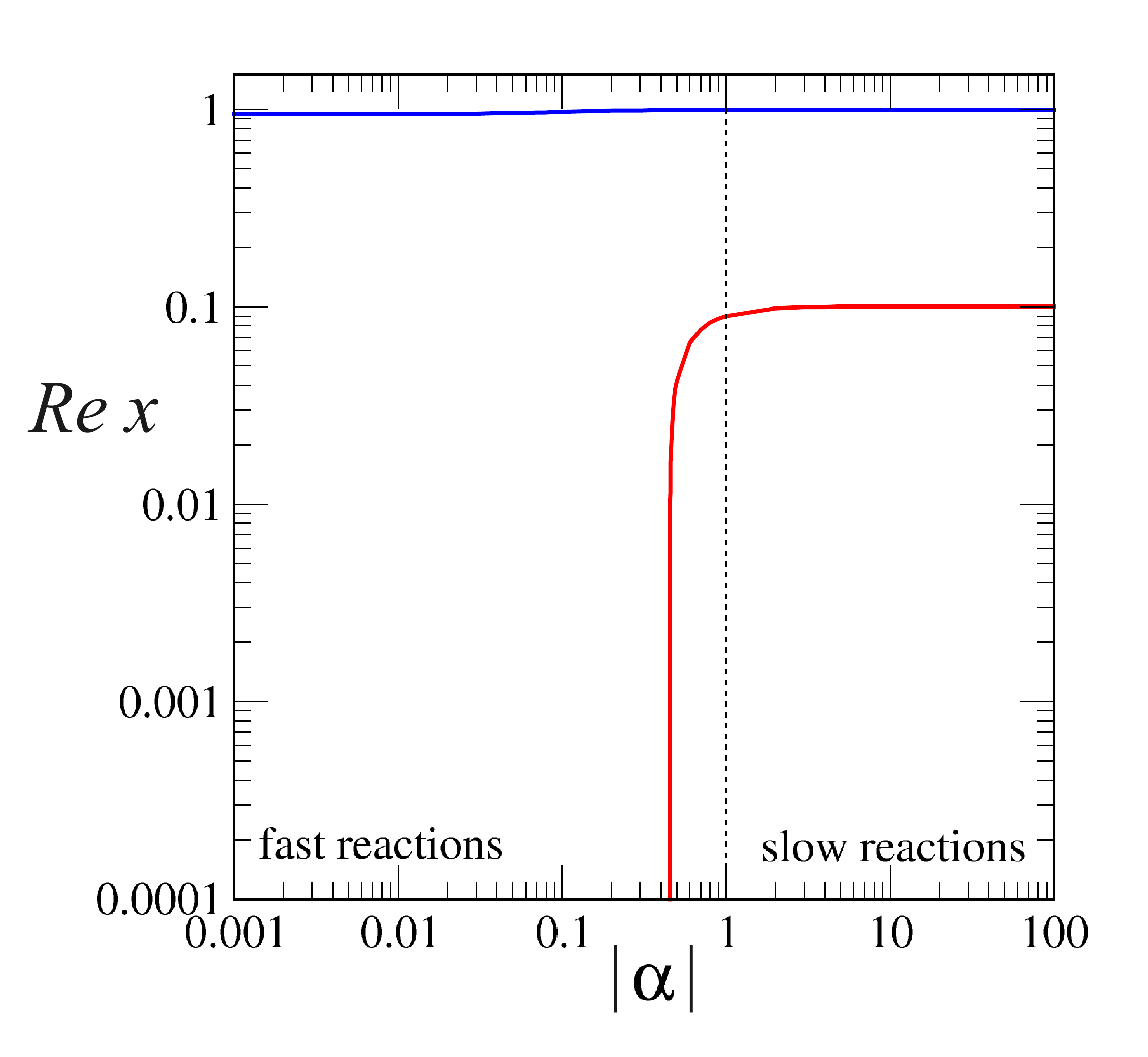}
\includegraphics[width=0.45\textwidth,clip]{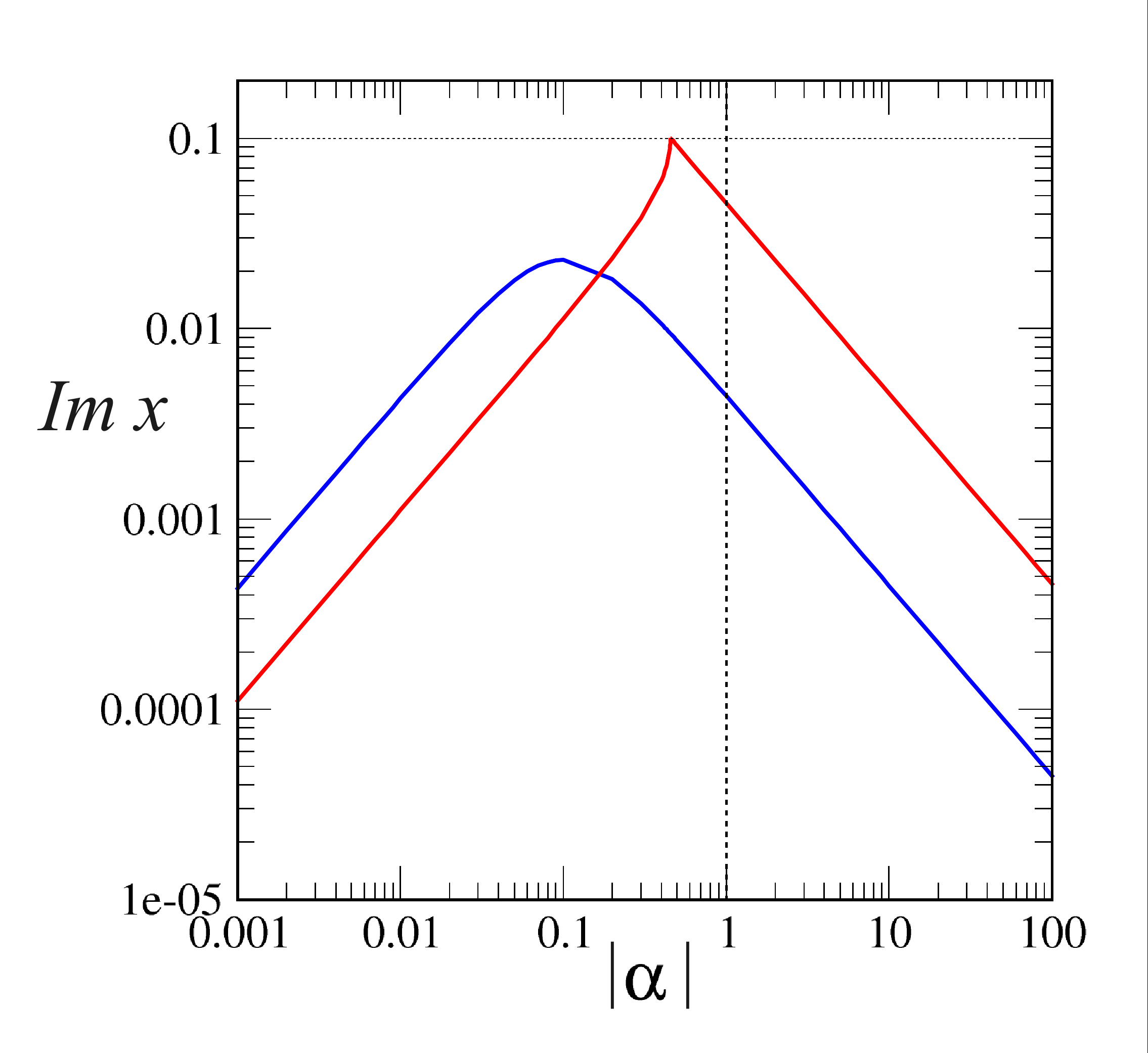}

\end{center}
\caption{An illustration of the transition from the fast to the slow reaction regime and the impact on the relevant roots to the polynomial \eqref{finpol}, representing acoustic modes (blue) and composition gravity modes (red). Top panel: We show the real parts of the mode frequencies for $\tilde \omega_0=0.95$ and $\tilde \omega_g = 0.1$ as functions of the (scaled) reaction time $\alpha$. Bottom panel: The corresponding imaginary parts. The vertical dashed line in each panel indicates where $|t_R \omega_g|=1$, the point at which the g-mode oscillation period is equal to the reaction rate. The results demonstrate the expected disappearance of the g-mode for  fast reaction rates.  The sharp disappearance of the oscillating g-mode (roughly)  when the imaginary part of the mode equals the real part (indicated by the horisontal dashed line) is notable.  The results  also bring out the expected resonance associated with the bulk viscosity damping of the acoustic mode.  } 
\label{fig1}
\end{figure}

The solutions to this problem illustrate the main principles, we are interested in. The top panel of figure~\ref{fig1} demonstrates the disappearance of the g-mode once the reaction timescale is shorter than that of the mode oscillation (when $|t_R \omega_g| \lesssim 1$). We also see that, at least in this model problem, the impact of the matter composition on the frequency of the sound waves is small. Meanwhile, the bottom panel of figure~\ref{fig1} illustrates how the damping of two  sets of modes changes as we vary the the reaction time. The sharp disappearance of the oscillating g-mode  when the imaginary part of the mode equals the real part is notable.   
%The results  also illustrate the expected resonance associated with the bulk viscosity damping of the f-mode.

\subsection{Recovering bulk viscosity}

At this point 
it is relevant to note that, in addition to demonstrating the disappearance of "slow" composition g-modes, our simple model problem highlights the resonant behaviour associated with bulk viscosity (apparent in the f-mode damping in figure~\ref{fig1}). This is not surprising since  reactions determine the rate of bulk viscosity in mature neutron stars [see, for example, \citet{alford}]. Nevertheless, our calculation provides an useful demonstration that the effect can be accounted for by explicitly allowing for the different particle fractions to evolve. In the extension, this approach could prove useful in nonlinear studies, where the deviation from chemical equilibrium is sufficiently large that the usual linearised approach cannot be applied. This may be particularly interesting given recent efforts to account for bulk viscosity in simulations of neutron star mergers \citep{bvsim} and the parallel development of a theoretical framework that would (at least in principle) allow us to account for non-conserved particle flows \citep{reactive}.

As it may be instructive to recover the standard prescription from the present formulation, note that the slow reaction expansion of \eqref{dbeta}
allows us to write \eqref{eqone} as
\begin{multline}
\Delta p \approx 
c_s^2  \Delta \rho +  \left( {\partial p \over \partial \beta} \right)_{\rho}  \mathcal B\left ( 1  - i { \mathcal A\over \omega} \right)\Delta \rho  \\
= \mathcal C  \Delta \rho + i { \mathcal A\over \omega} \left( {\partial p \over \partial \beta} \right)_{\rho}  \left( {\partial \beta \over \partial \rho}\right)_{x_\p} \Delta \rho \\
%=  \mathcal C  \Delta \rho - i { \rho  \mathcal A\over \omega} \left( {\partial p \over \partial \beta} \right)_{\rho}  \left( {\partial \beta \over \partial \rho}\right)_{x_\p} \nabla_j \xi^j
=   \mathcal C  \Delta \rho - \zeta  \nabla_j \xi^j
\end{multline}
where
\beq
\zeta =  i { \rho  \mathcal A\over \omega} \left( {\partial p \over \partial \beta} \right)_{\rho}  \left( {\partial \beta \over \partial \rho}\right)_{x_\p}
\eeq

%We now have
%\begin{multline}
%\delta p = \mathcal C \delta \rho + \left[ \mathcal C \xi^j \nabla_j \rho - \xi^j \nabla_j p \right]  - \zeta  \nabla_j \xi^j  \\
%=  \mathcal C \delta \rho -\rho   \mathcal C \left[ {1\over c_s^2} - {1\over \mathcal C}  \right] (\xi^j g_j)  - \zeta  \nabla_j \xi^j  \\
%=   \mathcal C \delta \rho - \rho \mathcal C (A_j \xi^j) - \zeta  \nabla_j \xi^j  
%\end{multline}

In order to use this result in the Euler equation, we may introduce
\beq
\delta p = \delta \bar p  - \zeta  \nabla_j \xi^j 
\eeq
such that $\delta \bar p$ represents the ``inviscid'' pressure perturbation. This immediately leads to 
\beq
-\omega^2  \xi_i +{1\over \rho}  \nabla_i \delta \bar p - {1\over \rho^2} \delta \rho \nabla_i p + \nabla_i \delta \Phi = \nabla_i (  \zeta  \nabla_j \xi^j )
\label{eulb2}
\eeq
where we recognize the term on the right-hand side as the bulk viscosity. 

%In essence, we should not be surprised that the model demonstrates the characteristic bulk viscosity ``resonance'' in the f-mode damping. In fact, this feature provides a useful sanity check of the %model.

\section{Implications}

We have demonstrated that nuclear reactions may remove composition g-modes from the oscillation spectrum of a mature neutron star.  This result should come as no surprise. The  transition from individual g-modes being present to them being absent may be a bit sharper than expected, but the overall behaviour is  intuitive. Nevertheless, there are valuable lessons to be learned.  In particular, one may want to be a bit careful with assumptions involving high-order (very low frequency) g-modes. Two topical scenarios  spring to mind: i) the saturation of modes driven unstable by gravitational-wave emission is thought to rely on the coupling to short-range (high overtone) modes \citep{pnig}, and ii) the tidal coupling to a pair of high overtone pressure p-modes and g-modes may lead to a non-resonant instability  \citep{2013ApJ...769..121W,pg2}. In both cases, the outcome may be affected (at least in principle) by the removal of g-modes from the low-frequency spectrum.

As a guide, let us consider the problem for a hot young neutron star for which the radiation driven instability may be particularly relevant\footnote{We will consider the tidal problem elsewhere, as it requires a more in depth discussion.}. To get a rough idea, we may use the estimate timescales from \citet{yak}. The relevant equilibration timescales are then\footnote{Given that we have assumed npe-matter we do not consider the, significantly faster, reactions associated with hyperons. Nevertheless, it is easy to see what the implications for that problem would be.}
$$
t_M \sim {2\ \mathrm{months} \over T_9^6} \ , \quad t_D \sim {20\ \mathrm{ s} \over T_9^4}
%\ , \quad t_H\sim {1\ \mathrm{ ms} \over T_9^2}
$$
for the modified and direct Urca reactions, respectively. The temperature is scaled to relatively hot systems, $T_9=T/10^9$~K. Taking $T_9=10$ for a typical proto-neutron star and assuming that g-modes disappear below a frequency 
\beq
\omega^\mathrm{cut} =2\pi f^\mathrm{cut} \approx {1\over t_R}
\eeq
in accord with the discussion in the previous section, we see that the spectrum would be altered below
$$
f^\mathrm{cut}_M \sim 3\times 10^{-2}~\mathrm{Hz} \ , \quad  f^\mathrm{cut}_D \sim 80~\mathrm{Hz}
%\ , \quad  f^\mathrm{cut}_H\sim 1.6\times 10^4~\mathrm{Hz} 
$$
in the two cases. Assuming that the g-modes reside at frequencies below (say) 100~Hz, we  see that only a few g-modes may be allowed in a hot star in which the direct Urca channel is open, with additional modes entering the spectrum as the star cools. In contrast, in the standard case of modified Urca reactions, there should be a large number of g-modes already at high temperatures, but very high overtone modes can still not exist. 

What do we learn from this? The conclusion may be more conceptual than of direct astrophysical relevance, but it is clear that one has to execute some level of care in discussions involving the dynamics of very high order g-modes. 
It is also relevant to consider how other aspects of neutron star physics enter the discussion. Superfluidity may be particularly important. After all, we know that superfluidity completely removes the g-modes for npe matter [as the superfluid neutrons may move relative to the charged components, see \citet{1995A&A...303..515L,comer}]. However, it is also known that the appearance of muons introduces relevant composition variation (now associated with the muon to electron ratio), which leads to the appearance of a set of (slightly higher frequency) g-modes \citep{kantor,pass}. These modes should, in principle, be affected by nuclear reactions. However, the neutron superfluidity also suppresses any reactions in which they are involved, which naturally affects the predicted cut-off frequency in the g-mode spectrum. Again, the outcome depends on the details. 

Finally, it is worth noting that our analysis was based on the assumption that the deviation from chemical equilibrium was sufficiently small that we could linearise the problem. This should be a valid assumption for many problems of interest, but one can easily think of situations where nonlinear aspects come into play (like neutron star mergers). It is well known that nonlinear  deviations from equilibrium  lead to shorter equilibration times \citep{hans,reis95,alford}. As this may have a significant effect on any estimated g-mode cut-off frequency, it is a problem worth further consideration. 

\section*{Acknowledgements}
Support from STFC via grant ST/R00045X/1 is gratefully acknowledged.

\bibliographystyle{mn2e}
%\bibliography{references}

\end{document}